\documentclass[aps, prd, showpacs, superscriptaddress, ctexart, nofootinbib,
twocolumn, preprintnumbers, floatfix, amsmath, amssymb,
superscriptaddress]{revtex4}

\usepackage{graphicx}
\usepackage{epsfig}
\usepackage{bm}
\usepackage{amsfonts} 
\usepackage{epstopdf}
\usepackage[latin1]{inputenc}
\usepackage{graphicx}
\usepackage{amssymb}
\usepackage{color}
\usepackage{float}
\usepackage{amsmath}
\usepackage{amsfonts}
\usepackage{dcolumn}
\usepackage{hyperref}
\usepackage{epsfig}
\usepackage{bm}

\newcommand{\beq}{\begin{equation}}
\newcommand{\eeq}{\end{equation}}
\newcommand{\bea}{\begin{eqnarray}}
\newcommand{\eea}{\end{eqnarray}}

\def\be{\begin{equation}}
\def\ee{\end{equation}}
\def\ba{\begin{eqnarray}}
\def\ea{\end{eqnarray}}

\def\be{\begin{equation}}
\def\ee{\end{equation}}
\def\ba{\begin{align}}
\def\ea{\end{align}}

\def\mc{\mathcal}
\def\t{\tilde}

\usepackage{color}

\begin{document}

\title{Cosmology of $F(R)$ nonlinear massive gravity}

\author{Yi-Fu Cai}
\email{yifucai@physics.mcgill.ca}
\affiliation{Department of Physics, McGill University, Montr\'eal, QC H3A
2T8, Canada}

\author{Emmanuel N. Saridakis}
\email{Emmanuel\_Saridakis@baylor.edu}
\affiliation{Physics Division, National Technical University of Athens, 15780
Zografou Campus,  Athens, Greece}
\affiliation{Instituto de F\'{\i}sica, Pontificia Universidad de Cat\'olica
de Valpara\'{\i}so, Casilla 4950, Valpara\'{\i}so, Chile}

\begin{abstract}
The theory of nonlinear massive gravity can be extended into the $F(R)$ form
as developed in Phys.Rev.D90, 064051 (2014). Being free of the Boulware-Deser ghost, such
a construction has the additional advantage of exhibiting no linear
instabilities around a cosmological background. We investigate various
cosmological evolutions of a universe governed by this generalized massive
gravitational theory. Specifically, under the Starobinsky ansantz, this model
provides a unified description of the cosmological history, from early-time
inflation to late-time self-acceleration. Moreover, under viable $F(R)$
forms, the scenario leads to a very interesting dark-energy phenomenology,
including the realization of the quintom scenario without any
pathology. Finally, we provide a detailed analysis of the cosmological
perturbations at linear order, as well as the Hamiltonian constraint
analysis, in order to examine the physical degrees of freedom.
\end{abstract}

\maketitle

\section{Introduction}

Einstein's General Relativity (GR) is commonly acknowledged as the standard
theory of gravitation at distance scales which are sufficiently large
compared to the Planck length. As a foundation of modern cosmology, GR has
been greatly successful in explaining the dynamics of the universe throughout
the whole thermal expanding history. Furthermore, recent high-precision
cosmological observations \cite{Ade:2013zuv} strongly support the standard
inflationary Big-Bang paradigm, in which our universe experienced a period of
inflation at early times, then evolved through radiation and matter
domination, and eventually entered a second cosmic acceleration today.
Although such a picture can be described in the GR framework, with some
additional scalar degrees of freedom, modifications of GR provide an
alternative explanation.

Amongst all modified gravity theories, the $F(R)$ paradigm is one simple
theory that can sufficiently describe the properties of higher-order
gravitational effects, by extending the gravitational Lagrangian as an
arbitrary function of the Ricci scalar
(see \cite{Sotiriou:2008rp,DeFelice:2010aj,Nojiri:2010wj} for
reviews). Hence, this
scenario is expected to have important effects at high-energy scales, that
were realized in the very early universe. In particular, the Starobinsky
model \cite{Starobinsky:1980te} is one of the earliest inflationary models,
and it lies in the center of the best-fit regime of the
recently released observational results by the  Planck collaboration
\cite{Ade:2013uln}. Moreover, the $F(R)$ modifications of GR can also be
applied to explain the cosmic acceleration of the late-time universe
\cite{Carroll:2003wy}, or inflationary and late-time epochs in a unified way 
\cite{Nojiri:2003ft}.

On the other hand, the question whether the graviton can acquire a mass, has
attracted the interest of theorists for decades. Initiated by Fierz and Pauli
\cite{Fierz:1939ix}, the subsequent full nonlinear formulation of massive
gravity was found to suffer from a severe problem, namely the existence of
the so-called Boulware-Deser (BD) ghost \cite{Boulware:1973my}. This
fundamental problem puzzled physicists until recently, where a specific
nonlinear extension of massive gravity was proposed by de Rham, Gabadadze and
Tolley (dRGT) \cite{deRham:2010ik}. Within this new theory, through a
Hamiltonian constraint analysis \cite{Hassan:2011hr} and an effective field
theory approach \cite{deRham:2011rn}, it is proven that the BD ghost can be
eliminated by a secondary Hamiltonian constraint (see
\cite{Hinterbichler:2011tt} for a review). Apart from the theoretical
interest, this nonlinear construction has the additional advantage that it
can provide an explanation of the observational evidence of the late-time
cosmic acceleration. In particular, the graviton potential can effectively
mimic a cosmological constant by fine tuning a sufficiently small value for
the graviton mass \cite{deRham:2010tw,Gumrukcuoglu:2011ew}. However, the
cosmological perturbations around these solutions exhibit in general severe
instabilities \cite{DeFelice:2012mx}.

Having the above in mind, it is interesting to consider the possibility of
combining the $F(R)$ paradigm and the recent nonlinear massive gravity, as
proposed in \cite{Cai:2013lqa}. Such an extension allows for a significantly
enhanced class of cosmological evolutions, especially describing inflation
and late-time acceleration in a unified way. However, the important
advantage is that this construction exhibits stable cosmological
perturbations at the linear level. In the present work we perform a detailed
analysis on the model of $F(R)$ massive gravity as an accompanied work to
\cite{Cai:2013lqa}. We first perform a detailed generalized Hamiltonian
constraint analysis in order to show the absence of the BD ghost. Moreover,
we investigate in detail various cosmological evolutions, from inflation to
dark energy eras, using viable $F(R)$ ansantzes and focusing on observables
such as the dark-energy equation-of-state parameter. Finally, we perform a
detailed analysis of the linear cosmological perturbations.

The paper is organized as follows. In Section \ref{intro} we briefly review
the paradigm of $F(R)$ nonlinear massive gravity. We then search for
cosmological solutions in both   Einstein and Jordan frames in Section
\ref{cosmologies}. In Section \ref{cosmoimpli} we present some examples of
the interplay between the $F(R)$ and the massive sectors of the gravitational
action, since this construction allows for a large variety of cosmological
evolutions in agreement with the observed behavior. In Section
\ref{cosmicpert} we perform a perturbation analysis at linear order around
the cosmological background, which  reveals only a single propagating scalar
mode. Since the kinetic term of the scalar graviton vanishes at leading order
the model is stable against linear perturbations Finally, we conclude with a
discussion in Section \ref{Sec:conclude}. Lastly, for completeness we present
the detailed analysis of the Hamiltonian constraint in the Appendix.

\section{$F(R)$ nonlinear massive gravity}\label{intro}

Following \cite{Cai:2013lqa}, the action of the $F(R)$ nonlinear massive
gravity is composed by the UV ($F(R)$ sector) and IR (``massive gravity''
sector) modifications, and it writes as
\begin{eqnarray}\label{action}
 S = \int d^4x \sqrt{-g} ~ M_p^2 \left[ \frac{1}{2}F(R) + m_g^2 \,{\cal
U}_{MG}\right] ~,
\end{eqnarray}
where $M_p=1/\sqrt{8\pi G}$ is the reduced Planck mass, $m_g$ the graviton
mass parameter, and $g_{\mu\nu}$ is the physical metric of the background
spacetime. As in the dRGT model, the dimensionless graviton potential ${\cal
U}_{MG}$ is constructed by a group of polynomials through the
anti-symmetrization in 4D spacetime as:
\begin{eqnarray}
\label{Utot}
 {\cal U}_{MG} = {\cal U}_2 + \alpha_3 {\cal U}_2 + \alpha_4 {\cal U}_2 ~,
\end{eqnarray}
with
\begin{eqnarray}
 &&{\cal U}_2 = \mathbb{K}^\mu_{[\mu} \mathbb{K}^\nu_{\nu]}
=\frac{1}{2}\left([\mathbb{K}]^2-[\mathbb{K}^2]\right)~, \nonumber\\
 &&{\cal U}_3 = \mathbb{K}^\mu_{[\mu} \mathbb{K}^\nu_{\nu}
\mathbb{K}^\sigma_{\sigma]} = \frac{1}{6}\left([\mathbb{K}]^3-3[\mathbb{K}]
[\mathbb{K}^2]+2[\mathbb{K}^3]\right) ~, \nonumber\\
 &&{\cal U}_4 = \mathbb{K}^\mu_{[\mu} \mathbb{K}^\nu_{\nu}
\mathbb{K}^\sigma_{\sigma} \mathbb{K}^\rho_{\rho]} = \frac{1}{24} \big(
[\mathbb{K}]^4-6[\mathbb{K}]^2[\mathbb{K}^2] +3[\mathbb{K}^2]^2  \nonumber\\
 && \ \ \ \ \ \ \ \ \ \ \ \ \  \ \ \ \ \ \ \ \ \ \ \ \ \ \ \
 +8[\mathbb{K}] [\mathbb{K}^3] -6[ \mathbb{K}^4] \big) ~.
\label{UMs}
\end{eqnarray}
In the above polynomials we have introduced the block matrix $\mathbb{K}
\equiv \mathbb{I} - \mathbb{X}$, where $\mathbb{X} \equiv \sqrt{g^{-1}f}$
involves a non-dynamical (fiducial) metric $f$. Additionally, $[\mathbb{K}]$
denotes the trace of the matrix $\mathbb{K}$. Finally,the coefficients
$\alpha_3$ and $\alpha_4$ appearing in the graviton potential are the two
model parameters.


Similarly to usual $F(R)$ gravity models, it is convenient to perform a
conformal transformation on the spacetime coordinate, from the original frame
(dubbed as  Jordan frame) to the frame where  gravity behaves like Einstein's
GR (called the Einstein frame). In particular, we impose the conformal
transformation as
\begin{eqnarray}\label{conformaltrans}
 g_{\mu\nu} \rightarrow \t{g}_{\mu\nu}=\Omega^2 g_{\mu\nu} ~,
\end{eqnarray}
with
$\Omega^2=F_{,R}=\exp\left[\sqrt{\frac{2}{3}}\frac{\varphi}{M_p}\right]$.
Correspondingly, the ``$F(R)$" sector of the gravitational Lagrangian can be
reformulated as
\begin{equation}
 \sqrt{-g} M_p^2\frac{F(R)}{2} \rightarrow \sqrt{-\t{g}} \big[
\frac{M_p^2}{2}\t{R} -\frac{1}{2} \t{g}^{\mu\nu}
\partial_\mu\varphi\partial_\nu\varphi -U(\varphi) \big] ~,
\end{equation}
where a potential for the scalar field is effectively introduced:
\begin{eqnarray}
 U(\varphi) = M_p^2 \bigg( \frac{RF_{,R}-F}{2 F_{,R}^2} \bigg) ~,
\end{eqnarray}
with the comma-subscript denoting differentiation with respect to the
following variable.

From the above formulation one could at first sight see a close relation
between the $F(R)$ massive gravity and the quasi-dilaton massive gravity
\cite{D'Amico:2012zv,Gannouji:2013rwa}. However, in the case of quasi-dilaton
massive gravity \cite{D'Amico:2012zv} the coefficient in front of the kinetic
term of the scalar field is a free parameter, while in our model it is set to
unity (this feature has a crucial effect on the perturbational analysis as we
will see). Moreover, the quasi-dilaton massive gravity is constructed upon a
shift symmetry along the scalar, while this symmetry is in general broken in
the present model due to the appearance of the effective potential
$U(\varphi)$. Thus, despite the similarity to the quasi-dilaton massive
gravity, the two models behave radically differently.

Additionally, the above conformal transformation acts on the gravitational
potential sector. Rewriting the graviton potential (\ref{Utot}) more
conveniently as
\begin{eqnarray}
\label{Utotnew}
 \mc{U}_{MG}=\sum_{i=0}^{4} \beta_i {\cal E}_i ~,
\end{eqnarray}
with
\begin{eqnarray}
 \beta_n=(-1)^i\big[\frac{1}{2}(4-i)(3-i)+(4-i)\alpha_3+\alpha_4 \big] ~,
\end{eqnarray}
and
\begin{align}
 & {\cal E}_0 = 1 ~,~~
 {\cal E}_1 = \mathbb{X}^\mu_\mu ~,~~
 {\cal E}_2 = \mathbb{X}^\mu_{[\mu } \mathbb{X}^\nu_{\nu ]} ~,~~\\
 & {\cal E}_3 = \mathbb{X}^\mu_{[\mu } \mathbb{X}^\rho_{\rho}
\mathbb{X}^\nu_{\nu]} ~,~~
 {\cal E}_4 = \mathbb{X}^\mu_{[\mu } \mathbb{X}^\rho_{\rho}
\mathbb{X}^\gamma_{\gamma} \mathbb{X}^\nu_{\nu ]} ~,
\end{align}
we deduce that in order to perform the conformal transformation
\eqref{conformaltrans}  we need to incorporate the transformation of the
matrix $\mathbb{X}\rightarrow\mathbb{\tilde X} \equiv
\sqrt{\tilde{g}^{-1}f}$. Since $\sqrt{-g} \rightarrow
\Omega^{-4}\sqrt{-\t{g}}$, we acquire
\begin{eqnarray}
 {\cal E}_i\rightarrow \Omega^i {\tilde{\cal E}}_i ~,
\end{eqnarray}
and finally we obtain the transformation as
\begin{align}
\label{potential}
 \mc{U}_{MG} \rightarrow \tilde{\cal U}_{MG} = \sum_{i=0}^4 \Omega^{i-4}
\beta_i {\cal E}_i ~.
\end{align}

From the expression (\ref{potential}) one can easily see that the
gravitational potential in the Einstein frame acquires a scalar-field
dependence. This feature is similar to the model of varying-mass massive
gravity \cite{Huang:2012pe, Saridakis:2012jy}. However, there exist two
differences. Firstly, in varying-mass massive gravity the graviton mass is an
arbitrary function of the scalar field, while in the present model the
scalar-field dependence is fixed by the conformal factor $\Omega$. Secondly
and more importantly, in the present model the power index of the
scalar-field-dependent (conformal) factor $\Omega$ is determined by the order
of the gravitational polynomial, while in varying-mass massive gravity it is
a common overall factor \cite{Huang:2012pe}. In other words, in the present
model the separate gravitational terms acquire a different scalar-field
dependence, which make the scenario radically different than that of
varying-mass massive gravity.

Eventually, to assemble the effects of both the ``$F(R)$'' and the ``massive
gravity'' sectors, we can rewrite the Lagrangian of the $F(R)$ nonlinear
massive gravity as:
\begin{equation}
\label{Einsteinframe}
 {\cal L} = \sqrt{-\tilde{g}} \bigg[ \frac{M_p^2}{2} ( \tilde{R} + 2m_g^2\,
\tilde{\cal U}_{MG} )
 -\frac{1}{2} \t{g}^{\mu\nu} \partial_\mu\varphi\partial_\nu\varphi -
U(\varphi) \bigg] .
\end{equation}
The scenario at hand is free of BD ghosts, as can be shown through a detailed
Hamiltonian constraint analysis, which for convenience is presented in the
Appendix.

\section{Cosmology of $F(R)$ nonlinear massive gravity}\label{cosmologies}

In this section we study the cosmological implications of the model
under investigation. We consider the standard homogeneous and isotropic
Friedmann-Robertson-Walker (FRW) ansatz for the physical metric, while for
the fiducial one we start from a Minkowski form
$f_{\rho\sigma}=\eta_{\rho\sigma}$. Furthermore, we incorporate the usual
matter content of the universe, that is we consider an action $S_m$
corresponding to an ideal fluid characterized by energy density $\rho_m$ and
pressure $p_m$ respectively, which  couples minimally to the gravitational
sector. Finally, we fix a particular gauge for the St\"uckelberg fields.

\subsection{Flat universe}
\label{flatgeneral}

We consider the physical metric in the Jordan frame to be flat FRW:
\begin{align}
\label{metric1}
 ds^2 = -N^2dt^2 +a^2\delta_{ij} dx^idx^j ~,
\end{align}
where we have omitted the subscript ``$g$'' to distinguish the physical
metric, since the fiducial metric is just the Minkowski one. Additionally, we
use part of the gauge freedom to impose the following form on the
St\"uckelberg fields:
\begin{align}
\label{stuckel1}
 \phi^0=b(t)~,~~\phi^i=a_c x^i ~,
\end{align}
with $a_c$ a constant. Inserting the above into the total action
\eqref{action} we obtain
\begin{align}
\label{actionflat}
 S_{tot} = &\int d^4 x \, a^3 N \bigg[\frac{M_p^2}{2}F(R) +m_g^2M_p^2
\big({\cal U}_2 +\alpha_3  {\cal U}_ 3 +\alpha_4{\cal U}_4 \big) \bigg]
\nonumber\\
 & + S_m ~,
\end{align}
with
\begin{align}
 {\cal U}_2 &= 3a(a-a_c)(2Na-a_cN-ab') ~, \nonumber\\
 {\cal U}_3 &= (a-a_c)^2(4Na-a_cN-3ab') ~, \nonumber\\
 {\cal U}_4 &= (a-a_c)^3(N-b') ~,
\end{align}
where the prime denotes the derivatives with respect to $t$.

For convenience, and in order to present the equations in a compact way, we
introduce the following notation:
\begin{align}
 &{\cal A}=\frac{a_c}{a} ~,~~ \dot{a}=\frac{a'}{N} ~,~~ H=\frac{\dot{a}}{a}
~,\nonumber\\
 &X_1=(3-2{\cal A})+\alpha_3(3-{\cal A})(1-{\cal A})+\alpha_4(1-{\cal A})^2
~,\nonumber\\
 &X_2=(3-{\cal A})+\alpha_3(1-{\cal A}) ~,
\end{align}
where $H$ is the usual Hubble parameter describing the expanding rate of the
universe. Varying the action (\ref{actionflat}) with respect to all field
variables, we obtain the following equations of motion:
\begin{align}
\label{flatJordan1}
 \delta_b S|_{N=1}: ~ & m_g^2M_p^2 H X_1=0 ~, \nonumber\\
 \delta_N S|_{N=1}: ~ & 3 M_p^{2} F_{,R} H^2 = \rho_m+\rho_{MG} +\rho_{F_R}
~, \nonumber\\
 \delta_a S|_{N=1}: ~ & M_p^{2} F_{,R} \left(-2\dot{H}-3H^2\right) =
p_m+p_{MG}+p_{F_R} ~,
\end{align}
where we have set $N=1$ at the end. In the above expressions we have defined
the effective energy density and   pressure of the ``massive gravity'' sector
as
\begin{align}
\label{rhomg}
 \rho_{MG} &=m_g^2M_p^2({\cal A}-1)(X_1+X_2)  ~, \\
\label{pmg}
 p_{MG} &=-m_g^2M_p^2({\cal A}-1)X_2-m_g^2M_p^2(\dot{b}-1)X_1 ~,
\end{align}
as well as the effective energy density and   pressure contributed by the
``$F(R)$'' sector as
\begin{align}
\label{rhofR}
 &\rho_{F_R} = M_p^2 \big[ \frac{R F_{,R}-F}{2}-3H\dot{R} F_{,RR} \big] ~,
\\
\label{pfR}
 & p_{F_R} = M_p^2 \big[ \dot{R}^2 F_{,RRR}+2H\dot{R} F_{,RR}
+\ddot{R}F_{,RR} +\frac{F-RF_{,R}}{2} \big] ~.
\end{align}
Here we would like to clarify that, if the model under consideration is used
to describe the cosmological dynamics at late times, the effective dark
energy component is attributed to the above two contributions, namely
\begin{align}
 &\rho_{DE}\equiv\rho_{MG}+\rho_{F_R} ~, \\
 &p_{DE}\equiv p_{MG}+p_{F_R} ~.
\end{align}

Solving the first equation of (\ref{flatJordan1}) leads to the following
possibilities: $H=0$, that is a static universe with $a(t)=const$; or
$X_1=0$, that is ${\cal A}=const$, which is a static universe too
($a(t)=const$). As a result, we find that the $F(R)$ nonlinear massive
gravity shares the disadvantage of many other nonlinear massive gravity
scenarios, namely that it does not accept flat homogeneous and isotropic
solutions (one can easily achieve the same conclusion in the Einstein frame).
Consequently, we have to extend the cosmological investigation of this model
into the non-flat geometry.

\subsection{Open Universe}

Since there exists no acceptable flat solution, in order to study
cosmological scenarios we consider an open universe. For completeness, we
analyze the cosmological dynamics in both Jordan and Einstein frames.

\subsubsection{Analysis in the Jordan frame}
\label{openJordan}

We start with the open FRW metric of the form
\begin{eqnarray}
\label{openmetric}
 ds^2=-N^2dt^2+a^2(t) \gamma^{K}_{ij}dx^idx^j ~,
\end{eqnarray}
with
\begin{eqnarray}
 \gamma^{K}_{ij} dx^idx^j = \delta_{ij}dx^idx^j -
\frac{a_k^2(\delta_{ij}x^idx^j)^2}{1-a_k^2\delta_{ij}x^ix^j} ~,
\end{eqnarray}
and $a_k=\sqrt{|K|}$ is associated with the spatial curvature. The
St\"uckelberg scalars take the form of the Milne coordinates:
\begin{align}
 \varphi^0 = b(t) \sqrt{ 1 + a_k^2 \delta_{ij}x^ix^j } ~,~~ \varphi^i = a_k
b(t) x^i ~.
\end{align}
We insert the above formulae into the total action \eqref{action} and we
obtain
\begin{eqnarray}
 &&S_{tot} = \int d^4 x\sqrt{\gamma^K}a^3 N \left[\frac{M_p^2}{2}F(R)
\right.\nonumber\\
&&\left.\ \ \ \ \ \  +m_g^2M_p^2({\cal U}_2 +\alpha_3 {\cal U}_3+\alpha_4
{\cal U}_4)\right]+S_m ,
\label{actionopenJ}
\end{eqnarray}
with
\begin{align}
 {\cal U}_2 &=3a(a-a_k b)(2{N}{a}-{b'}{a}-{N}a_k b)~,\nonumber\\
 {\cal U}_3 &=({a}-a_k b)^2(4{N}{a}-3{a}{b'}-Na_k  b)~,\nonumber\\
 {\cal U}_4 &=({a}-a_k b)^3({N}-{b'})~,
\end{align}
where primes denote derivatives with respect to $t$ as introduced in the
previous subsection.

Similar to the analysis in the flat universe we introduce a series of
notation for convenience, namely
\begin{align}
 &{\cal B}=\frac{a_kb}{a} ~,~~ \dot{a}=\frac{a'}{N} ~,~~ H=\frac{\dot{a}}{a}
~, \nonumber\\
 &Y_1=(3-2{\cal B})+\alpha_3(3-{\cal B})(1-{\cal B})+\alpha_4(1-{\cal B})^2
~, \nonumber\\
 &Y_2=(3-{\cal B})+\alpha_3(1-{\cal B}) ~.
\end{align}
Then varying the action (\ref{actionopenJ}) with respect to the field
variables yields the following equations of motion:
{\small{
\begin{align}
\label{openJordan1}
 \delta_b S|_{N=1}: ~ &(\dot{a}-a_k)Y_1 = 0 ~,  \\
\label{openJordan2}
 \delta_N S|_{N=1}: ~ &3 M_p^{2} F_{,R} \left( H^2-\frac{a_k^2}{a^2}
\right) = \rho_m+\rho_{MG}+\rho_{F_R}, \\
\label{openJordan3}
 \delta_a S|_{N=1}: ~ &M_p^{2} F_{,R}\left(-2\dot{H}-3H^2
+\frac{a_k^2}{a^2}\right) = p_m +p_{MG} +p_{F_R} .
\end{align}}}
In the above expressions the effective energy density and pressure of the
``$F(R)$'' sector remain the same as in the flat geometry, which are given by
(\ref{rhofR}) and (\ref{pfR}), while the effective energy density and
pressure of the ``massive gravity'' sector now become
\begin{align}
\label{rhomgjordan}
 \rho_{MG} &= m_g^2M_p^2({\cal B}-1)(Y_1+Y_2) ~,\\
\label{pmgjordan}
 p_{MG} &= -m_g^2M_p^2\left[({\cal B}-1)Y_2+(\dot{b}-1)Y_1\right] ~.
\end{align}
If this model is used to describe late-time cosmology, the effective dark
energy component is attributed to the above two contributions through
$\rho_{DE}\equiv\rho_{MG} +\rho_{F_R}$ and $p_{DE} \equiv p_{MG}+p_{F_R}$.

Let us now examine the cosmological equations
(\ref{openJordan1})-(\ref{openJordan3}). Similar to all massive gravity
scenarios, Eq. \eqref{openJordan1} constrains the dynamics significantly. As
we observe it leads to two possible solutions. The first one is trivial:
$\dot{a}=a_k$, which leads to the dynamics $a(t)=a_k t+const$. The second one
requires $Y_1=0$, and is of physical interest. As pointed out in the
self-accelerating backgrounds of the dRGT construction
\cite{Gumrukcuoglu:2011ew}, the nontrivial constraint $Y_1=0$ yields
\begin{align}
 {\cal{B}}_\pm = \frac{1 +2\alpha_3 +\alpha_4\pm\sqrt{1 +\alpha_3 +\alpha_3^2
-\alpha_4}}{\alpha_3 +\alpha_4} ~.
\label{Bpm}
\end{align}
This relation can be always fulfilled by choosing $b(t)\propto a(t)$. In this
case expressions (\ref{rhomgjordan}), (\ref{pmgjordan}) imply that
$\rho_{MG}=-p_{MG}=const$, as it is expected similarly to the original
nonlinear massive gravity model. This result incorporates the interesting
cosmological implication that the graviton mass can induce an effective
cosmological constant given by (\ref{rhomgjordan}), of which the energy
density takes the form
\begin{equation}
\label{effcc}
 \rho_{MG\pm} = m_g^2 M_p^2 ({\cal B}_\pm-1) \left[(3-{\cal
B}_\pm)+\alpha_3(1-{\cal B}_\pm) \right] ~.
\end{equation}
However, the crucial issue is that in the present model the remaining
``$F(R)$" sector can be chosen at will in the Friedmann equations
(\ref{openJordan2})-(\ref{openJordan3}), leading to a huge class of
cosmologies.

\subsubsection{Analysis in the Einstein frame}
\label{openEinst}

For completeness, we present the investigation of the open geometry in the
Einstein frame too. We mention that this will be helpful in the
examination of the perturbations of the scenario, since the analysis of
cosmological perturbations takes a familiar form in the Einstein frame. We
follow the procedure of section \ref{intro}. Specifically, we perform the
conformal transformation $g_{\mu\nu} \rightarrow \t{g}_{\mu\nu}=\Omega^2
g_{\mu\nu}$ of the open FRW metric (\ref{openmetric}), with $\Omega^2 =
\exp\left[\sqrt{\frac{2}{3}}\frac{\varphi}{M_p}\right]$. Therefore, the
metric in the Einstein frame is given by
\begin{eqnarray}
\label{openmetric}
 && d\t{s}^2 = \Omega^2 \left[-N^2dt^2+a^2(t) \gamma^{K}_{ij}dx^idx^j\right]
\nonumber\\
 &&\ \ \ \ \ =-\t{N}^2d\t{t}^2+\t{a}^2(\t{t}) \gamma^{K}_{ij}dx^idx^j ~,
\end{eqnarray}
and the resulting action in the Einstein frame takes the form
\begin{eqnarray}
\label{actionopenE}
 &&S = \int d^4x \sqrt{\gamma^{K}} \left[ -3M_p^2a_k^2
\t{N}\t{a}-3M_p^2\frac{\t{a} \t{a}'^2}{\t{N}}\ \ \ \ \ \ \ \ \ \,
\right.\nonumber\\
&&\ \ \ \ \  \ \ \ \ \ \ \ \ \ \ \ \ \ \ \ \ \
+m_g^2M_p^2(\tilde{\cal U}_2 +\alpha_3\tilde{\cal U}_3 +\alpha_4\tilde{\cal
U}_4)\nonumber\\
&&\ \ \ \ \ \ \ \ \ \ \ \ \ \ \  \  \ \ \ \ \
-\frac{\t{a}^3\t{N}}{2} \partial^\alpha\varphi\partial_\alpha\varphi
-\t{N}\t{a}^3U(\varphi)\nonumber\\
&&\left.\ \ \ \ \ \ \ \ \ \ \ \ \ \ \  \  \ \ \ \ \
+\t{N}\t{a}^3\t{\mathcal{L}}_m\right],
\end{eqnarray}
with $\t{\mathcal{L}}_m=\Omega^{-4} \mathcal{L}_m$ describing any additional
matter and
\begin{align}
 \tilde{\cal U}_2 &= 3\t{a}\Omega^{-4}(\t{a}-a_k\Omega b)(2\t{N}\t{a}-\Omega
b'\t{a}-\Omega\t{N}a_k b ) ~, \nonumber\\
 \tilde{\cal U}_3 &= \Omega^{-4}(\t{a}-a_kb\Omega)^2(4\t{N}\t{a}-3\t{a}\Omega
b' -Na_k\Omega b) ~, \nonumber\\
 \tilde{\cal U}_4 &= \Omega^{-4}(\t{a}-a_kb\Omega)^3(\t{N}-\Omega b') ~,
\end{align}
where primes denote derivatives with respect to $\t{t}$. Note that, as we
mentioned in section \ref{intro}, the conformal transformation acts only on
the physical metric and not on the fiducial one, and therefore the
St\"uckelberg fields remain unaffected.

We introduce a series of notation for simplicity, namely
\begin{align}
\label{Einstein_notation}
 \tilde{\cal B} &= \frac{a_k b\Omega}{\t{a}} ~,~~
 \dot{\t{a}}=\frac{\t{a}'}{\t{N}} ~,~~ \t{H}=\frac{\dot{\t{a}}}{\t{a}} ~,
\nonumber\\
 \t{Y}_1 &= (3-2\tilde{\cal B}) +\alpha_3(3-\tilde{\cal B})(1-\tilde{\cal B})
 +\alpha_4(1-\tilde{\cal B})^2 ~, \nonumber\\
 \t{Y}_2 &= (3-\tilde{\cal B}) +\alpha_3(1-\tilde{\cal B}) ~.
\end{align}
The equations of motion in this case are obtained by
\begin{align}
\label{OpenEOMb}
 \delta_b S|_{\t{N}=1}: ~ &\Big(a_k
 -\dot{\t{a}}+\t{a}\frac{\dot{\Omega}}{\Omega}\Big)\t{Y}_1 =0 ~, \\
\label{OpenEOMN}
 \delta_{\t{N}}
 S|_{\t{N}=1}: ~
 &3M_p^2\Big(\t{H}^2-\frac{a_k^2}{\t{a}^2}\Big)=\t{\rho}_m+\t{\rho}_{MG}
 +\t{\rho}_ {\varphi} ~,\\
 \delta_{\t{a}} S|_{\t{N}=1}: ~ &M_p^{2} \left(-2\dot{\t{H}}-3\t{H}^2
 +\frac{a_k^2}{\t{a}^2}\right) =\t{p}_m+\t{p}_{MG} \nonumber\\
   &  \ \ \ \ \ \ \  \ \ \ \  \ \  \ \ \ \ \ \ \  \ \ \ \  \ \
 \ \ \ \ \ \ \  \ \ \,  +\t{p}_ {\varphi} ~,\\
\label{OpenEOMO}
 \delta_\varphi S|_{\t{N}=1}: ~ &\ddot{\varphi}+3\t{H}\dot{\varphi}
 +U_{,\varphi} -\frac{m_g^2M_p^2}{\Omega^5} \Omega_{,\varphi}
 \left[3\t{Y}_1 \Omega \dot{b}\right. \nonumber\\
 &\left.+4(1-\tilde{\cal B})\t{Y}_2
 +(4-\tilde{\cal B}) \t{Y}_1 \right]\nonumber\\
 &-\frac{\Omega_{,\varphi}}{\Omega}\left(-\t{\rho}_m+3\t{p}_m \right)=0,
\end{align}
where we have set $\t{N}=1$ in the final expressions, and in the last
equation used the relation \cite{DeFelice:2010aj}
\begin{eqnarray}
\frac{\partial\t{\mathcal{L}}_m}{\partial\varphi}
=\sqrt{-\t{g}}\frac{\Omega_{,\varphi}}{\Omega}\left(-\t{\rho}_m+3\t{p}
_m\right) ~.
\end{eqnarray}
As usual, in order to express the results back to the initial metric we use
\begin{eqnarray}
\label{tildett}
 d\t{t}=\Omega dt ~,~~ \t{a}=\Omega a ~,~~
 \t{H}=\frac{1}{\Omega}\left(H+\frac{\dot{\Omega}}{\Omega}\right) ~.
\end{eqnarray}
In the above expressions, the effective energy density and pressure of the
regular matter component are given by
\begin{align}
 \t{\rho}_m = \frac{\rho_m}{\Omega^4} ~,~~ \t{p}_m = \frac{p_m}{\Omega^4} ~,
\end{align}
and those of the ``massive gravity" sector are expressed as
\begin{align}
\label{rhomgeinstein}
 \t{\rho}_{MG} &= \frac{m_g^2M_p^2}{\Omega^4} (\tilde{\cal B}-1)
(\t{Y}_1+\t{Y}_2) ~, \\
\label{pmgeinstein}
 \t{p}_{MG}&=-\frac{m_g^2M_p^2}{\Omega^4} \left[ (\tilde{\cal B}-1) \t{Y}_2
+(\Omega \dot{b}-1)\t{Y}_1 \right] ~.
\end{align}
Additionally, there exists one more contribution to the total energy density
and pressure, namely that of the scalar field, which take the usual forms
\begin{align}
 \t{\rho}_ {\varphi} = \frac{1}{2}\dot{\varphi}^2+U(\varphi) ~,~~
\t{p}_{\varphi} = \frac{1}{2}\dot{\varphi}^2-U(\varphi) ~.
\end{align}
Finally, in the analysis of the Einstein frame of $F(R)$ nonlinear massive
gravity, the corresponding dark energy component is attributed to the
``massive gravity'' and the scalar field sectors, namely
\begin{align}
 \t{\rho}_{DE} = \t{\rho}_{MG}+\t{\rho}_ {\varphi} ~,~~ \t{p}_{DE} \equiv
\t{p}_{MG}+\t{p}_ {\varphi} ~.
\end{align}

Now let us examine the cosmological equations
(\ref{OpenEOMb})-(\ref{OpenEOMO}). Similarly to the analysis of the Jordan
frame, Eq. (\ref{OpenEOMb}) leads to two possible solutions. The first is the
trivial solution $\t{H} = \frac{a_k}{\t{a}} +\frac{\dot{\Omega}}{\Omega}$,
which using (\ref{tildett}) is found to be the Einstein-frame equivalent of
the trivial solution $\dot{a}=a_k$ found in the Jordan frame. The second is
$\t{Y}_1=0$, which further yields
\begin{align}
 \tilde{{\cal{B}}}_\pm = \frac{1 +2\alpha_3 +\alpha_4\pm\sqrt{1 +\alpha_3
+\alpha_3^2 -\alpha_4}}{\alpha_3 +\alpha_4} ~.
\end{align}
This relation can be always fulfilled by choosing $b(\t{t}) \propto
{a}(\t{t})$. Then expressions (\ref{rhomgeinstein}), (\ref{pmgeinstein})
imply that $\t{\rho}_{MG}=-\t{p}_{MG}$, as we found in the Jordan frame too.
Note that in the Einstein frame $\t{\rho}_{MG}$ and $\t{p}_{MG}$ depend on
$\t{t}$ (through the conformal factor $\Omega$), but if we re-express them
back into the Jordan frame, in terms of $t$ they become constant. Hence, one
can easily deduce that this interesting time varying cosmological constant in
the Einstein frame, is due to the conformal factor brought by the coordinate
transformation.

\section{Phenomenological Implications}\label{cosmoimpli}

Having presented the basic background equations of motion of the scenario of
$F(R)$ nonlinear massive gravity,  we are now able to investigate its
phenomenological implications.  We mention that, the scenario at hand
exhibits a wide class of cosmological behaviors due to the combination of the
``$F(R)$'' and the ``massive gravity'' sectors \footnote{A similar analysis
within the frame of $F(R)$ bi-gravity was performed in
\cite{Nojiri:2012zu}.}.

As we discussed in the previous section, the cosmological equations in an
open FRW universe governed by $F(R)$ nonlinear massive gravity are given by
(\ref{openJordan1})-(\ref{openJordan3}). The first of these equations
constrains the dynamics significantly, leading to a constant contribution of
the graviton mass sector, that is $\rho_{MG\pm}=const$, given by
(\ref{effcc}). However, the crucial issue is that in the model at hand, and
contrary to usual massive gravity, the remaining ``$F(R)$" sector can be
chosen at will in the  rest Friedmann equations
(\ref{openJordan2})-(\ref{openJordan3}), leading to very rich cosmological
dynamics.

\subsection{The Starobinsky-$\Lambda$CDM-like cosmology}
\label{Sec:Starobinsky}

\begin{figure}[ht]
\includegraphics[scale=0.4]{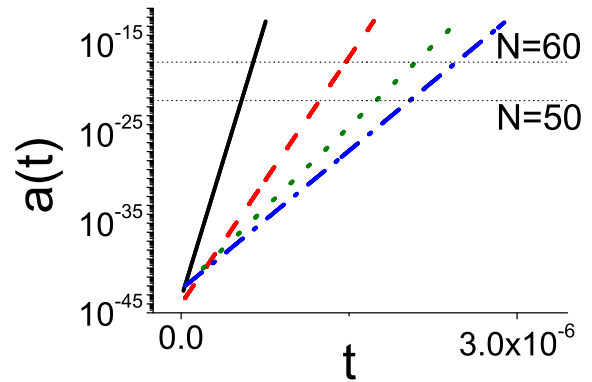}
\caption{
{\it{Four inflationary solutions for the Starobinsky ansatz (\ref{Sraro1}),
corresponding to
a) $m_g=10^{-50}$, $\alpha_3=2$, $\alpha_4=-1$, $a_k=5\times10^{-41}$,
$\xi=10^{10}$ (black-solid),
b) $m_g=10^{-50}$, $\alpha_3=10$, $\alpha_4=10$, $a_k=10^{-41}$,
$\xi=10^{9}$ (red-dashed),
c) $m_g=10^{-50}$, $\alpha_3=1$, $\alpha_4=1$, $a_k=10^{-40}$,
$\xi=10^{10}$ (blue-dash-doted),
d) $m_g=10^{-50}$, $\alpha_3=10$, $\alpha_4=1$, $a_k=2\times10^{-41}$,
$\xi=10^{10}$ (green-dotted).
All parameters are in Planck units. The two horizontal lines mark the $N=50$
and $N=60$ e-folding regimes.}} }
\label{Inflation.Staro}
\end{figure}

One of the most important and interesting sub-classes is when the ``$F(R)$"
sector is important at early times and thus responsible for inflation, while
the massive graviton is dominant at late times and can drive the universe
acceleration as observed today. To give a representative example, we
particularly consider the well-known Starobinsky's $R^2$-inflation scenario
\cite{Starobinsky:1980te}
\begin{eqnarray}
\label{Sraro1}
 F(R) = R+ \frac{\xi}{M_p^2} R^2 ~,
\end{eqnarray}
with $\xi$ being the high-order coupling coefficient. This model proves to be
the best-fitted scenario with the recently released Planck data
\cite{Ade:2013uln}. We perform a numerical elaboration of such a cosmological
system and in Figure \ref{Inflation.Staro} we present the early-time
inflationary solutions for four parameter choices.
\begin{figure}
\includegraphics[scale=0.4]{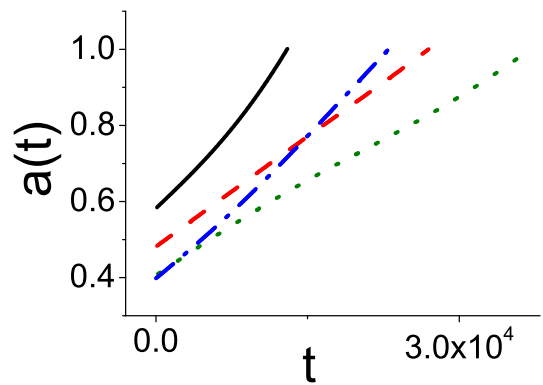}
\caption{
{\it{Four late-time accelerating solutions  for the Starobinsky ansatz
(\ref{Sraro1}), corresponding to
a) $m_g=2$, $\alpha_3=1$, $\alpha_4=1$, $a_k=0.05$,
$\xi=0.2$ (black-solid),
b) $m_g=1.5$, $\alpha_3=1$, $\alpha_4=-2$, $a_k=0.01$, $\xi=0.5$
(red-dashed),
c) $m_g=3$, $\alpha_3=3$, $\alpha_4=-5$, $a_k=0.05$, $\xi=0.5$
(blue-dash-doted),
d) $m_g=3$, $\alpha_3=10$, $\alpha_4=1$, $a_k=0.05$, $\xi=0.5$
(green-doted).
All parameters are in units where the present Hubble parameter is $H_0=1$,
and we have imposed $\Omega_{m0}\approx0.31$, $\Omega_{DE0}\approx0.69$,
$\Omega_{k0}\approx0.01$ at the present scale factor $a_0=1$
\cite{Ade:2013zuv}. }}
}
\label{Darkenergy.Staro}
\end{figure}

Additionally, in Figure \ref{Darkenergy.Staro} we numerically depict four
late-time self-accelerating solutions. We would like to mention that in order
to handle the late-time evolutions, we need to define the dark energy
equation of state $w_{DE}$ and its density parameter $\Omega_{DE}$. In the
$F(R)$-literature there are two ways to do it. The first is to use the
Friedmann equations in the form of (\ref{openJordan2}), (\ref{openJordan3}),
in which we introduce an effective Planck mass square $M_p^2F_{,R}$, and
define $\Omega_{F_R}\equiv\rho_{F_R}/(3M_p^2F_{,R}H^2)$, with $w_{F_R}\equiv
p_{F_R}/\rho_{F_R}$ \cite{Sotiriou:2008rp}. The second method is to rewrite
the Friedmann equations in order to have the standard Planck mass square
$M_p^2$ in the left hand side, to absorb the remaining terms in a modified
$\bar{\rho}_{F_R}$ and $\bar{p}_{F_R}$, and to define
$\Omega_{F_R}\equiv\bar{\rho}_{F_R}/(3M_p^2H^2)$, with $w_{F_R} \equiv
\bar{p}_{F_R}/\bar{\rho}_{F_R}$ \cite{DeFelice:2010aj}. Despite the fact
that
in usual cases the difference is small at the background level (and the
specific
choice has to be made in accordance with the specific measurement method of
$M_p$, $\Omega_{DE}$ and $w_{DE}$), we stress that the usual conservation
equation $\dot{\rho}+3H(\rho+p)=0$ holds only for $\bar{\rho}_{F_R}$ and not
for $\rho_{F_R}$ \cite{DeFelice:2010aj}. Although, this is not crucial (since
we are dealing with modified gravity the dark energy sector is effective,
arising from the extra gravitational features, and there is no fundamental
reason that it should be conserved), in the present work we follow the second
way in order for our model to present energy conservation of the ``$F(R)$"
sector, and subsequently for the whole dark-energy one. Thus, the matter,
curvature and massive-gravity density parameters are defined using
$3M_p^2H^2$ too, and they are also conserved.

As we observe from these figures, in this particular choice of the
Starobinsky ansatz, the Ricci scalar is relative large at early times, and
thus the $R^2$ correction term drives a successful inflation. On the other
hand, $R$ becomes very small at late times and thus the ``$F(R)$"
contribution is dramatically suppressed, and therefore the acceleration is
driven solely
by the effective cosmological constant induced by the graviton mass.
Therefore, this model provides a unified description of both inflation and
late-time acceleration of the universe. In particular, the corresponding
cosmological evolution at the background level provides a specific
realization of the Starobinsky-$\Lambda$CDM cosmology. Note however that at
the perturbation level the present model will behave differently comparing to
Starobinsky-$\Lambda$CDM cosmology, since the graviton sector contributes as
well, contrary to the simple cosmological constant which does not.
Therefore, strictly speaking, it is a  Starobinsky-$\Lambda$CDM-like
scenario.

\subsection{Late-time Cosmology}

As we mentioned above, in the case where the universe evolves at late times,
both the ``$F(R)$" and ``massive gravity" sectors may contribute effectively
to the dark energy component. In particular, the ``$F(R)$" contribution
brings dynamical features into dark energy physics, and thus it can be of
great phenomenological interest. Thus, in this subsection, we focus on the
details of late-time evolutions, going beyond the simple effective
cosmological constant of the previous subsection, and using viable  $F(R)$
forms, which quantitatively describe the observational data.
\begin{figure}
\includegraphics[scale=0.4]{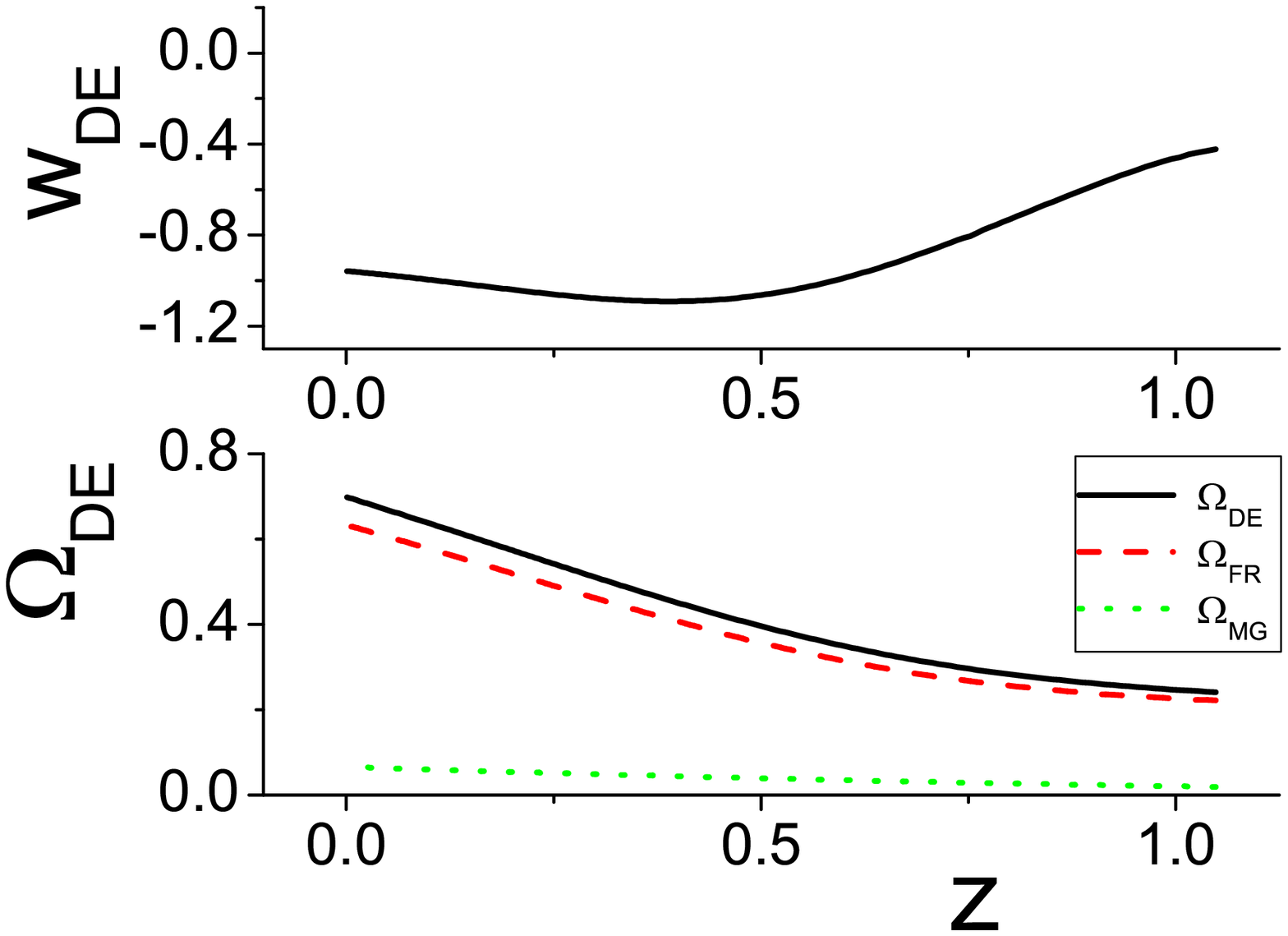}
\caption{
{\it{Upper graph: The dark energy equation-of-state parameter $w_{DE}$ as a
function of the redshift $z$, in the case of the exponential $F(R)$ ansatz
(\ref{expansatz}), for $m_g=1 $, $\alpha_3=3$, $\alpha_4=-2$,
$a_k=0.1$,
$\beta=1.5$, $R_S=0.25$.  All parameters are in units where the present
Hubble parameter is $H_0=1$,
and we have imposed  $\Omega_{m0}\approx0.31$, $\Omega_{DE0}\approx0.69$ and
$\Omega_{k0}\approx0.01$ at present \cite{Ade:2013zuv}. Lower graph: The
corresponding evolution of the dark energy density parameter $\Omega_{DE}$
(back-solid), as well as it  two constituents $\Omega_{F_R}$ (red-dashed) and
$\Omega_{MG}$ (green-dotted), of the $F(R)$ and massive-gravity sectors
respectively.   }}
}
\label{exponential1}
\end{figure}
\begin{figure}[!]
\includegraphics[scale=0.4]{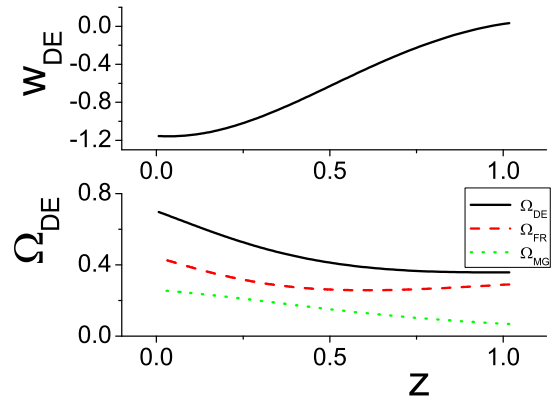}
\caption{
{\it{Upper graph: The dark energy equation-of-state parameter $w_{DE}$ as a
function of the redshift $z$, in the case of the exponential $F(R)$ ansatz
(\ref{expansatz}), for $m_g=0.75 $, $\alpha_3=1$, $\alpha_4=-2$,
$a_k=0.1$,
$\beta=1.1$, $R_S=0.3$. All parameters are in units where the present
Hubble parameter is $H_0=1$,
and we have imposed $\Omega_{m0}\approx0.31$, $\Omega_{DE0}\approx0.69$ and
$\Omega_{k0}\approx0.01$ at present \cite{Ade:2013zuv}. Lower graph: The
corresponding evolution of the dark energy density parameter $\Omega_{DE}$
(back-solid), as well as its two constituents $\Omega_{F_R}$ (red-dashed) and
$\Omega_{MG}$ (green-dotted), of the $F(R)$ and massive-gravity sectors
respectively. }}
}
\label{exponential2}
\end{figure}

We would like to examine the two well-known viable $F(R)$ models that are in
best agreement with observations. First we consider the $F(R)$ gravity of
the exponential ansatz \cite{Zhang:2005vt,Li:2006vi,Elizalde:2010ts}
\begin{equation}
\label{expansatz}
 F(R)=R-\beta R_{S}\left(1-e^{-R/R_{S}}\right) ~,
\end{equation}
with $R_{S}$ and $\beta$ the model parameters. This ansatz is viable for
$\beta>1$ and $R_{S}>0$ \cite{Zhang:2005vt}, and it is able to fit
observations with only one more parameter comparing to the $\Lambda$CDM
cosmology. We numerically solve the cosmological system governed by the
$F(R)$ massive gravity with this exponential form and we present two
different results in Figures \ref{exponential1} and \ref{exponential2}, under
two different groups of parameter choices, respectively. Our results focus on
the dynamics of the equation-of-state parameter of dark energy $w_{DE}$ and
the corresponding density parameter $\Omega_{DE}$. For clarity, we also
depict the evolutions of both dark energy constituents, namely the separate
$F(R)$ density parameter $\Omega_{F_R}$ and the massive gravity one
$\Omega_{MG}$, where $\Omega_{DE} = \Omega_{F_R} +\Omega_{MG}$. For
convenience, we use the redshift $z=\frac{a_0}{a}-1$ as the variable of the
horizontal axis, with the present scale factor $a_0$ set to $1$.

In the case of Figure \ref{exponential1} the dark energy is mainly induced by
the ``$F(R)$" sector, since the ``massive gravity" contribution is quite
small. On the other hand, in the case of Figure \ref{exponential2}, both the
``$F(R)$" and ``massive gravity" sectors have a significant contribution to
$\Omega_{DE}$. One notices that in these specific examples, the equation of
state parameter $w_{DE}$ evolves from $w>-1$ into the $w<-1$ regime and hence
provides a realization of the quintom scenario. This is known as a capability
and advantage of the $F(R)$ gravity models \cite{Nojiri:2006ri}.

Concerning the phantom-divide crossing realization (see \cite{Cai:2009zp} for
a comprehensive review), there is a well-known ``No-Go" theorem which states
that any dynamical dark energy with only one single degree of freedom (DoF)
in the frame of General Relativity cannot exhibit it, due to severe gradient
instabilities in the dispersion relation at the classical level
\cite{Xia:2007km}. In general, in order to circumvent this problems, the
quintom models usually involve two scalar DoFs
\cite{Feng:2004ad}, or they are formulated in alternative frameworks such is
the non-perturbative string-inspired construction \cite{Cai:2007gs}, or
using a spinor field \cite{Cai:2008gk}. Interestingly, in the present case of
$F(R)$ nonlinear massive gravity, although at linear level we have only one
scalar DoF as we will show in section \ref{cosmicpert}, its interaction with
the graviton potential modifies the dispersion relation which becomes
well-behaved when the phantom-divide crossing occurs. Thus, the
aforementioned ``No-Go" theorem is avoided (this mechanism is similar to the
``No-Go" theorem avoidance in the Horndeski models \cite{Horndeski:1974wa,
Deffayet:2011gz}).

\begin{figure}
\includegraphics[scale=0.4]{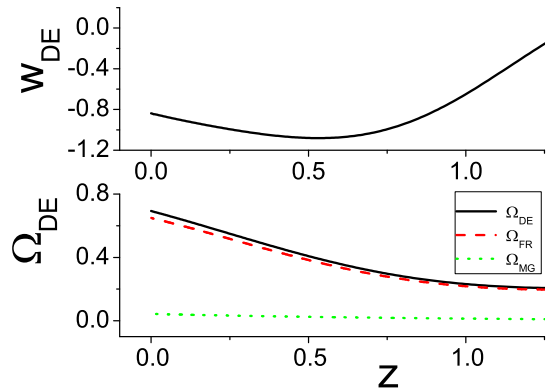}
\caption{
{\it{Upper graph: The dark energy equation-of-state parameter $w_{DE}$ as a
function of the redshift $z$, in the case of the power-law-like Starobinsky
$F(R)$ ansatz (\ref{powerStarobinsky}), for  $m_g=0.5 $, $\alpha_3=2$,
$\alpha_4=1$, $a_k=0.1$, $\lambda=0.3$, $n=2$, $R_c=2$. All parameters are in
units where the present
Hubble parameter is $H_0=1$,
and we have imposed  $\Omega_{m0}\approx0.31$,
$\Omega_{DE0}\approx0.69$ and $\Omega_{k0}\approx0.01$ at present
\cite{Ade:2013zuv}. Lower graph: The corresponding evolution of the dark
energy density parameter $\Omega_{DE}$ (back-solid), as well as its two
constituents $\Omega_{F_R}$ (red-dashed) and $\Omega_{MG}$ (green-dotted), of
the $F(R)$ and massive-gravity sectors respectively.}}
}
\label{power1}
\end{figure}

\begin{figure}
\includegraphics[scale=0.4]{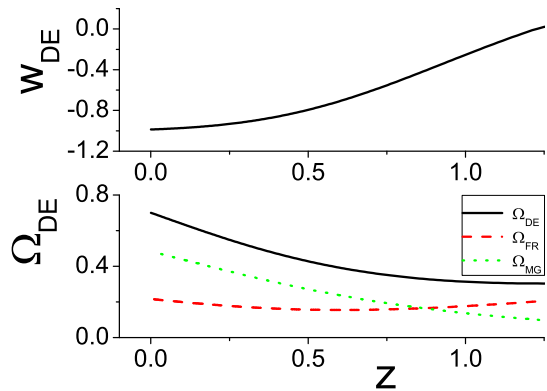}
\caption{
{\it{Upper graph: The dark energy equation-of-state parameter $w_{DE}$ as a
function of the redshift $z$, in the case of the power-law-like Starobinsky
$F(R)$ ansatz (\ref{powerStarobinsky}), for $m_g=2 $,
$\alpha_3=2$, $\alpha_4=1$, $a_k=0.1$, $\lambda=25$, $n=2$,
$R_c=10$.  All parameters are in units where the present
Hubble parameter is $H_0=1$,
and we have imposed  $\Omega_{m0}\approx0.31$, $\Omega_{DE0}\approx0.69$ and
$\Omega_{k0}\approx0.01$ at present \cite{Ade:2013zuv}. Lower graph: The
corresponding evolution of the dark energy density parameter $\Omega_{DE}$
(back-solid), as well as its two constituents $\Omega_{F_R}$ (red-dashed) and
$\Omega_{MG}$ (green-dotted), of the $F(R)$ and massive gravity
sectors, respectively.}}
}
\label{power2}
\end{figure}

The second viable and well-studied $F(R)$ ansatz is of the power-law-like
Starobinsky form \cite{Starobinsky:2007hu, Capozziello:2007eu}:
\begin{equation}
\label{powerStarobinsky}
 F(R) = R-\lambda R_{c} \bigg[ 1 -\left(1+\frac{R^{2}}{R_{c}^{2}}\right)^{-n}
\bigg] ~,
\end{equation}
with $\lambda$, $n$, and $R_{c}$ the (positive) model parameters ($n>0.9$
according to Solar-System constraints \cite{Capozziello:2007eu}). We
numerically solve the cosmological equations under the above $F(R)$ form and
we provide two representative results in Figures \ref{power1} and
\ref{power2}, respectively.

In the case shown in Figure \ref{power1}, the dynamics of the dark energy is
dominated by the contribution of the ``$F(R)$" sector, since the ``massive
gravity" contribution is very small. From the evolution of $w_{DE}$ in this
figure, one can see that the quintom scenario occurs twice, thus the universe
remains in the phantom regime for a short time. This interesting
phenomenological property can be applied in the study of very early universe
and a nonsingular bouncing solution can be obtained, similarly to
\cite{Cai:2007qw}. On the other hand, in the case shown in Figure
\ref{power2}, the ``massive gravity" sector has the largest contribution at
late times, comparing to the $F(R)$ part. Note that the domination of the
``massive gravity" sector in Figure \ref{power2}   makes $w_{DE}$ to approach
$-1$ but does not allow for a crossing. This is expected since, as we
mentioned after (\ref{Bpm}), the pure ``massive gravity" sector behaves like
a cosmological constant (in the case of asymptotic domination of
$\Omega_{MG}$ on $\Omega_{F_R}$, the $w_{DE}$ tends asymptotically to $-1$).

\section{Cosmological Perturbations}\label{cosmicpert}

The scenario of $F(R)$ nonlinear massive gravity is free of the BD ghost, as
it is shown in detail in the Appendix. However, it is necessary to examine
whether this model remain free of instabilities at the perturbative level.
The stability issue of various nonlinear massive gravity models has recently
been extensively studied in the literature \cite{Gumrukcuoglu:2011zh,
DeFelice:2012mx, Gumrukcuoglu:2013nza, Khosravi:2012rk}, and the parameter
spaces of these models are tightly constrained.

In this section, we proceed to a detailed study of the cosmological
perturbations of the scenario at hand, focusing our interest at the level of
linear perturbations, and this analysis serves as a completeness of our
accompanied work \cite{Cai:2013lqa}. We work in the unitary gauge which
leaves the St\"uckelberg fields unperturbed. Additionally, as usual we
perform the calculations in the Einstein frame, but we drop the tildes on the
metric for convenience (thus in this section the dots denote derivatives with
respect to the time-variable of the Einstein-frame metric).

The metric variables found in the action, and introduced in Eq.
(\ref{openmetric}), are perturbed according to the following convention
\begin{align}
 &g_{00} = -N^2(1+2\phi)~,\nonumber\\
 &g_{0i} =Na\beta_i ~,\nonumber\\
 &g_{ij}=a^2(\gamma^K_{ij}+h_{ij}) .
\end{align}
The scalar field is also perturbed as $\varphi\rightarrow
\varphi+\delta\varphi$ while the St\"uckelberg fields are not as we work in
the unitary gauge. Therefore, note that in this section the symbol $\varphi$
is used for the scalar field in the Einstein frame, while $\phi$ is used for
the scalar metric perturbation (thus no confusion is possible since the
St\"uckelberg fields do not appear).

Then we can further decompose them in a traceless and transverse tensor
($\gamma_{ij}$), transverse vectors ($B_i,~F_i$) and scalars
($\phi,~B,~\psi,~E$) parts through the following relations,
\begin{align}
 &\beta_i =\partial_i B + B_i ~,~~D_i B^i=0,\\
 &h_{ij} =2\gamma^K_{ij}\psi+ D_iD_j
E+\frac{1}{2}(D_iF_j+D_jF_i)+\gamma_{ij},\\
 & D_iF^i =D^i\gamma_{ij}=0 ~,~~ \gamma^i_i=0~,
\end{align}
where we have introduced the spatial covariant derivative $D_i$ compatible
with the background 3-metric $\gamma^K_{ij}$.

One can derive the equations of motion of cosmological perturbation by
perturbing the extended Einstein's equation of nonlinear massive gravity. We
do not include the contribution from extra matter as it does not change the
conclusion. Einstein's equation can be expressed as
\begin{align}
 \delta G_{\mu\nu}+\delta X_{\mu\nu}=M_p^{-2}\delta T_{\mu\nu} ~,
\end{align}
where $G_{\mu\nu}$ is the regular Einstein tensor and the extra term
$X_{\mu\nu}$ comes from the ``massive gravity" part of the action. For
convenience we introduce
\begin{eqnarray}
 &&M_{GW} =\frac{m_g^2 \tilde{\cal B}}{8}(1-\tilde{\cal B})
\left[ 1 +(2-\tilde{\cal B})\alpha_3 +(1-\tilde{\cal B})\alpha_4 \right]
  \nonumber\\
 &&\ \ \ \ \ \ \ \ \ \  \ \ +\frac{m_g^2 \tilde{\cal B}}{8}\t{Y}_1 ~,
\end{eqnarray}
which will be used in the following calculation.

The process of deriving the
perturbed Einstein equations is straightforward. Here we list the
non-vanishing equations of motion as follows,
{\small{
\begin{align}
\label{00comp}
 &
 3H(H\phi-\dot{\psi})
+\frac{D^2}{a^2}\left[\psi+aH\left(B-\frac{a\dot{E}}{2}\right)\right]
+\frac{3a_k^2\psi}{a^2} =
\nonumber\\
&-\frac{\dot{\varphi}\dot{\delta\varphi}+U_{,\varphi}\delta\varphi}
{ 2 } +\frac{\dot{\varphi}^2\phi}{2} +
\frac{2m_g^2M_p^2\Omega_{,\varphi}}{\Omega^5} (\tilde{\cal B}-1)
\t{Y}_2\delta\varphi ~,
\\
\label{0icomp}
 & \left[ H\phi -\dot\psi -\frac{a_k}{a}\left(B-\frac{a\dot{E}}{2}\right)
\right]_{,i} = \left[\frac{\dot{\varphi} \delta\varphi}{2} \right]_{,i} ~, \\
\label{iicomp}
 & \  -(3H^2+\dot{H})\phi -H\dot{\phi} +\ddot{\psi} +3H\dot{\psi}
-\frac{a_k^2}{a^2} (\phi+\psi) \nonumber\\
&\
-\frac{D^2}{3} \left[ \frac{\psi+\phi}{a^2}
+\frac{2H}{a}\left(B-\frac{a\dot{E}}{2}\right)
+\frac{1}{a}\frac{ d }{ dt }\left(B-\frac{a\dot{E}}{2}\right)
\right] \nonumber\\
 &\  +M_{GW} \psi\nonumber\\
 & \  =
-\frac{\dot{\varphi}\dot{\delta\varphi}-U_{,\varphi}\delta\varphi}{2}
-\frac{2m_g^2M_p^2\Omega_{,\varphi}}{\Omega^5} (\tilde{\cal
B}-1)\t{Y}_2\delta\varphi ~, \\
\label{ijcomp}
 & \left(\frac{D^2+3a_k^2}{a^2}\right) \left[
\frac{2H}{a}\left(B-\frac{a\dot{E}}{2}\right) +\frac{1}{a}\frac{ d }{ dt
}\left(B-\frac{a\dot{E}}{2}\right) \right.\nonumber\\
&\left. \ \ \ \ \ \ \ \ \ \ \ \ \ \ \ \ \ \ \
+\phi+\psi -\frac{M_{GW}}{2}E \right] =0
~,
\end{align}}}
which correspond to the $(00)$, $(0i)$, $(ii)$, $(ij)$ components,
respectively. Additionally, we have the perturbed Klein-Gordon equation for
the field fluctuation, which takes the form of
\begin{align}
\label{scalareom}
 &\ddot{\delta\varphi} +3H\dot{\delta\varphi} -\frac{D^2}{a^2}\delta\varphi
+U_{,\varphi\varphi}\delta\varphi +2U_{,\varphi}\psi
-4\dot{\varphi}\dot{\psi} = \nonumber\\
 & \  M_{GW}\left(\frac{\dot{\varphi}}{2}\dot{E} -U_{,\varphi}E\right)
+8m_g^2\Omega^{-4} (\tilde{\cal B}-1) \t{Y}_2 \delta\varphi ~,
\end{align}
where the coefficient $\tilde{\cal B}$ is given by \eqref{Einstein_notation}.

Since neither $\phi$ nor $\dot\psi$ appear as coefficient of the graviton
mass term, the kinetic structure is very similar to that of GR. This is seen
in many other massive gravity constructions, when one considers the
self-accelerating cosmological solution as the background. With this in mind,
we are allowed to use the Bardeen potentials,
\begin{align}\label{bardeenp}
 & \psi_B = -\psi -aH \big( B-\frac{a\dot{E}}{2} \big) ~,\nonumber\\
 & \phi_B = \phi +\frac{d}{dt} \bigg[ a \big(B-\frac{a\dot{E}}{2} \big)
\bigg]~,\nonumber\\
 & \delta\varphi_B = \delta\varphi +\dot{\varphi} a\big(B-\frac{a\dot{E}}{2}
\big)~,
\end{align}
to rewrite the equations and reduce the apparent propagating DoFs in the
present model. Making use of the above varialbe, Eq. \eqref{ijcomp} yields
\begin{eqnarray}
\label{potentialconstraint}
 \phi_B-\psi_B=M_{GW} \frac{E}{2} ~.
\end{eqnarray}

Similar to usual perturbation analysis, it is convenient to define the
well-known Mukhanov-Sasaki variable:
\begin{eqnarray}
 {\cal{Q}}=\delta\varphi_B+\frac{\dot{\varphi}}{H}\psi_B ~.
\end{eqnarray}
We first combine the $(00)$ and $(ii)$ components of the perturbed Einstein
equations, by calculating Eq. \eqref{00comp} minus \eqref{iicomp}, and then
insert the result into the perturbed Klein-Gordon equation \eqref{scalareom}.
In terms of the variable ${\cal{Q}}$, we obtain the perturbation equation as
follows:
{\small{
\begin{align}
\label{eom_Q}
 & \ddot{\mathcal{Q}} +3H\dot{\cal{Q}} +\left[\frac{D^2}{a^2}
+U_{,\varphi\varphi} -\frac{1}{M_p^2a^3} \frac{ d }{ dt } \left(
\frac{a^3}{H}\dot{\varphi}^2 \right) \right] {\cal{Q}} = \nonumber\\
 &\frac{2 m_g^2\tilde{Y}_Q}{3\Omega^4} {\cal{Q}} -\frac{2a_k^2}{a^2H^2} \Big(
\ddot{\varphi}-\frac{\dot{H}\dot{\varphi}}{H} \Big) \psi_B ~,
\end{align}}}
where $\tilde{Y}_Q\equiv 4(1-\tilde{\cal B}) \tilde{Y}_2$.

Observing Eq. \eqref{eom_Q}, one deduces that in the scenario at hand the
field variable ${\cal{Q}}$ is one propagating DoF characterizing the dynamics
of cosmological perturbations. The crucial issue is to examine whether
$\psi_B$ is a dynamical DoF too, or whether it can be integrated out through
some constraint equation.

We first re-express equations \eqref{00comp} and \eqref{0icomp} in terms of
the Bardeen potentials \eqref{bardeenp} and ${\cal{Q}}$. Therefore, we can
derive the following two useful formulae:
\begin{align}
\label{psi_B}
 \psi_B
 &= C^{-1}\left(\frac{\ddot{\varphi}}{2M_p^2}{\cal{Q}}
+\frac{\dot{\varphi}^3}{4M_p^4H}{\cal{Q}}
-\frac{\dot{\varphi}}{2M_p^2}{\dot{\cal{Q}}} \right) ~, \\
 \dot{\psi}_B
 &= C^{-1}\left(-\frac{\ddot{\varphi}\dot{\varphi}^2}{4M_p^4 H}
-\frac{\ddot{\varphi}H}{2M_p^2} +\frac{\dot{H}\dot{\varphi}^3}{4M_p^4 H^2}
\right.\nonumber\\
&\left.
\ \ \ \ \ \ \ \ \ \ \ \  -\frac{\dot{\varphi}^3}{M_p^4}
-\frac{3\dot{H}\dot{\varphi}}{2M_p^2}
-\frac{D^2}{a^2}\frac{\dot{\varphi}}{2M_p^2} \right){\cal{Q}} \nonumber\\
 &
 \ \ \ +C^{-1}\left(\frac{\dot{\varphi}^3}{4M_p^4 H} +\frac{\dot{\varphi}
H}{2M_p^2} \right) {\dot{\cal{Q}}} \nonumber\\
 &
 \ \ \
 +C^{-1}\left(\frac{D^2}{a^2} +3H\dot{H}
+\frac{3H\dot{\varphi}^2}{2M_p^2}\right.\nonumber\\
&\left.
\ \ \ \ \ \ \ \ \ \ \ \ \ \ \ -\frac{\dot{H}\dot{\varphi}^2}{2M_p^2 H}
-\frac{\dot{\varphi}^4}{2M_p^2 H} \right)M_{GW} \frac{E}{2} ~,
\label{dotpsi_B}
\end{align}
with the operator $C$   defined as
\begin{eqnarray}
 C =  -\frac{D^2}{a^2} -3\dot{H} -\frac{3\dot{\varphi}}{2M_p^2}
+\frac{\dot{H}\dot{\varphi}^2}{2M_p^2 H^2} +\frac{\dot{\varphi}^4}{4M_p^4
H^2} ~.
\end{eqnarray}
Equations (\ref{psi_B}) and (\ref{dotpsi_B}) determine the dynamics of
$\psi_B$. However, they are not independent since
$\frac{ d }{ dt }\psi_B=\dot{\psi}_B$. The consistency of their combination,
together with the main perturbation equation \eqref{eom_Q}, leads to a strong
constraint, namely $E=0$. Hence, we can clearly see that there exists only a
single scalar propagating DoF, that is ${\cal{Q}}$, at the linear order in
the perturbation theory around the cosmological background.

Let us now examine the stability of the single DoF ${\cal{Q}}$. Inserting
\eqref{psi_B} into \eqref{eom_Q}, we observe that the sign in front of
$\ddot{\cal{Q}}$ remains equal to one, and thus no apparent ghost instability
appears. The existence of the term containing $\psi_B$ would modify the
dispersion relation of the gradient terms. But in a realistic cosmological
background   the spatial curvature term is observationally suppressed and
thus the corresponding effect on modifying the dispersion relation would be
secondary. The last interesting property shown in Equation \eqref{eom_Q} is
the term proportional to $m_g^2$. Obviously, this is a new term brought by
the gravitational potential and thus it is interpreted as the effective mass
of the scalar graviton. Despite   the corrections from other terms, one
deduces that if $\tilde{Y}_Q<0$ then the model is free of tachyonic
instabilities. Note that this requirement still leaves a large part of the
parameter space, and thus a huge class of interesting
cosmological behavior.

We close this section by making some comments on the comparison with other
extended nonlinear massive gravity models, such are the quasi-dilaton massive
gravity \cite{D'Amico:2012zv} and the mass-varying massive gravity
\cite{Huang:2012pe}, following the discussion of section \ref{intro}. From
the point of view of perturbation analysis, both the quasi-dilaton and
mass-varying massive gravity models involve at least two scalar DoF, one
introduced by the additional scalar field and the other being the
longitudinal mode of the graviton. This can be verified by counting the
number of nonzero eigenvalues of the matrix for the kinetic part of the
perturbation action \cite{Gumrukcuoglu:2013nza}. Applying the method of
\cite{Gumrukcuoglu:2013nza} in the present scenario of $F(R)$ nonlinear
massive gravity (this can be performed immediately by taking the results of
\cite{Gumrukcuoglu:2013nza} and set the coefficient in front of the
scalar-field kinetic term to unity), we find that there exists only one
nonzero eigenvalue in our model, and this implies only a single DoF. Thus,
the present model is radically different from the aforementioned two.

\section{Conclusions}
\label{Sec:conclude}

Recently, a nonlinear massive gravity theory was constructed in
\cite{deRham:2010ik}, in which the BD ghost can be removed in
the decoupling limit to all orders in perturbation theory, through a
systematic construction of a covariant nonlinear action. The theoretical
advantages of such a construction led to a wide investigation of its
cosmological implications  \cite{deRham:2010tw, Gumrukcuoglu:2011ew}, as well
as of the black hole  solutions \cite{Koyama:2011xz}, while the connections
to bi-metric and multi-metric gravity were also revealed \cite{Damour:2002wu,
Hassan:2011tf}. However, the cosmological perturbations around these
solutions are found to exhibit in general severe instabilities
\cite{DeFelice:2012mx}, and thus extensions of the basic theory are
necessary.

In this work, we constructed the $F(R)$ extension of nonlinear massive
gravity, as an accompanied paper to our previous work \cite{Cai:2013lqa},
focusing on a detailed cosmological investigation, both at the background and
perturbation levels. In particular, we investigated three representative
cosmological evolutions. In the first one we considered the $F(R)$ sector to
be of the Starobinsky's $R^2$-form and we explicitly showed that this model
can describe both inflation and late time acceleration in a unified picture,
with the $F(R)$ sector driving inflation and the graviton mass responsible
for dark energy. Furthermore, we studied the case in which $F(R)$ takes the
usual viable exponential form and we showed that this class of
models can easily quantitatively describe the late-time universe behavior,
including the possibility of a phantom-divide crossing. In the last example
we considered the power-law-like Starobinsky $F(R)$ ansantz, which is another
viable form, with similar quantitative behavior. In these two classes of
evolutions, the effective dark energy component includes contributions from
both sectors, namely the $F(R)$ and the massive graviton one.

Apart from the interesting cosmological implications at the background level,
the advantage of the present scenario is the behavior of cosmological
perturbations. We developed the perturbations by expanding the Einstein
equations to linear order, around a cosmological solution. Although the
dispersion relation of the scalar mode is different from the case of GR, the
effect is negligible in a realistic cosmological background due to the
suppression of the curvature term. Our analysis showed that there exists only
one propagating scalar mode in this model, which can additionally be free of
ghost instabilities in a large part of the parameter space. This issue is the
crucial difference of $F(R)$ nonlinear massive gravity comparing to other
extensions, such are the quasi-dilaton and the varying-mass massive gravity.
The fact that both $F(R)$ and massive graviton sectors are needed for this
behavior, may imply that the UV and IR modifications of gravity may not be
independent.

At the end of the present work, we would like to highlight one interesting
property of the specific model of $R^2$-massive gravity. One notices that,
the effective potential for the scalar field after conformal transformation,
approaches a constant in the UV regime. This extremely flat potential, which
can be applied to drive sufficiently long inflationary process at early
universe as analyzed in Sec. \ref{Sec:Starobinsky}, indicates an
approximately shift symmetry along the scalar field. This feature
interestingly coincides with the model of quasi-dilaton massive gravity with
a specifically chosen dilatonic parameter. However, when the scalar field
evolves into the IR regime, it is stabilized at the vacuum and hence this
shift symmetry can be spontaneously broken. This profound property is
interesting to the process of symmetry breaking in nonlinear massive gravity
models, and deserves further investigations.

\begin{acknowledgments}
We are grateful to R. Brandenberger, P. Chen, A. De Felice, S. Deser,  G.
Gabadadze, A. E. G\"umr\"uk\c{c}\"uo\u{g}lu, C.~-C.~Lee, C. Lin, S.
Mukohyama, M. Salgado, T. Sotiriou, M. Trodden and S. Tsujikawa for useful
discussions. We also very much acknowledge to F. Duplessis for initial 
collaboration and extensive discussion in this project. The work of CYF 
is supported in part by the Department of Physics in McGill University. 
The research of ENS is implemented within the framework
of the Operational Program ``Education and Lifelong Learning'' (Actions
Beneficiary: General Secretariat for Research and Technology), and is
co-financedby the European Social Fund (ESF) and the Greek State. ENS wishes
to thank the Department of Physics of McGill University, for the hospitality
during the preparation of this work.
\end{acknowledgments}

\appendix

\section{Hamiltonian Constraint Analysis}\label{hamconst}

In this appendix we perform a Hamiltonian constraint analysis of $F(R)$
nonlinear massive gravity, and we show that the potentially dangerous extra
mode of graviton scalar vanishes due to the existence of a secondary
constraint. We follow the method developed in \cite{Hassan:2011hr} (see also
\cite{Cai:2013lqa,Kluson:2013yaa}), and we reproduce the major derivation for
completeness. As usual, we perform the analysis in the  Einstein frame
based on the Lagrangian \eqref{Einsteinframe}, and we drop the
tilde-subscripts for convenience.

Using the ADM formalism one can decompose the physical and fiducial metric as
{\small{
\begin{align}
 & \tilde{g}_{\mu\nu} dx^\mu dx^\nu = -(N_g^0)^2dt^2+\gamma_{ij}
(dx^i+N_g^idt) (dx^j+N_g^jdt) , \nonumber\\
 & f_{\sigma\rho} dx^\sigma dx^\rho = -(N_f^0)^2dt^2+\omega_{ab}
(dx^a+N_f^adt) (dx^b+N_f^bdt) .
\end{align}}}
Amongst all the coefficients in the metrics, the lapse $N_g^0$ and the shift
$\vec{N}_g$ (the three $N_g^i$'s expressed as vector) of the physical metric,
as well as the corresponding ones for the fiducial metric, $N_f^0$ and
$\vec{N}_f$ respectively, are not dynamical DoFs.

In general massive gravity $\gamma_{ij}$ allows for at most six propagating
modes, one of them being the origin of the BD ghost. Thus, a potentially
healthy theory must maintain a single constraint on ${\bar{\gamma}}$ (from
now on a bar denotes the  matrix form) and the corresponding conjugate
momenta, along with an associated secondary constraint, which will lead to
the elimination of the ghost DoF. In the following we show the existence of
these constraints in $F(R)$ massive gravity.

In general, the propagating DoFs are the $\{\gamma_{ij},\varphi\}$, such that
their phase space spans $14$ dimensions after including their conjugate
momentas $\{\pi_{ij},\pi\}$. In standard GR, four ghostly propagating DoF are
eliminated by the constraints from the equations of motion of the lapse $N$
and shifts $N^i$. The inclusion of a generic mass term changes these
equations and they do not directly give rise to any constraints on
$\gamma_{ij}$. However, the dRGT mass term is such that the
four equations of motion for $N$ and $N^i$ give rise to an equation
independent of $N$ and $N^i$, which provides a constraint
that eliminates one potentially dangerous DoF. Consistency of the first
constraint to be valid at later times introduces a secondary constraint.
These are necessary since we expect massive gravity to have only five
propagating DoF carried by $\gamma_{ij}$. The extra $6$-th mode carried by
the helicity $0$ mode is at the origin of the BD ghost.

We can make this constraint explicit by introducing a Lagrange multiplier in
the action. This is done by using a new shift function $n^i$ defined through
the following relation:
\begin{align}
 N^i-L^i=(L\delta^i_j+ND^i_j)n^j,
\end{align}
where $D^i_j$ is determined by
\begin{align}
\label{Dexp}
 \sqrt{(1-n^i\xi_{ij}n^j)}D^i_j=\sqrt{(\gamma^{ik}-D^i_mn^mD^k_ln^l)\xi_{kj}}
\nonumber .
\end{align}
The explicit form of $D$ is not needed for the analysis, but without loss of
generality we apply the expression that satisfies the identity $\xi D = D
\xi$. Defining $\lambda=(1-n^i\xi_{ij}n^j)$, the Hamiltonian can be written
as
\begin{equation}
\label{hamil}
 H=\int d^3x \mc{H}=\int d^3x (\mc{H}_0+N\mc{C}) ~,
\end{equation}
with
\begin{align}
\label{H_blocks}
 \mc{H}_0 &= -(Ln^i+L^i)\mc{R}_i-L\sqrt{\gamma}\mc{A} m_g^2 M_p^2
\nonumber\\
 \mc{C} &= \mc{R}+\mc{R}_iD^i_j n^j+\sqrt{\gamma}\mc{B}m_g^2 M_p^2 ~,
\nonumber
\end{align}
where
\begin{align}
 &\mc{R} = \frac{\sqrt{\gamma}}{2} \left[M_p^2 \text{}^{(3)}R
-\partial_i\varphi\partial^i\varphi -2U(\varphi) \right. \nonumber\\
 &\left.\ \ \ \ \ \ +\frac{1}{\gamma} \left(\frac{1}{M_p^2}\pi^i_i\pi^j_j
-\frac{2}{M_p^2}\pi^{ij}\pi_{ij} -\frac{1}{2}\pi^2\right) \right] ~,
\nonumber\\
 &\mc{R}_i = 2\gamma_{ij}\nabla_k\pi^{jk} -\pi\partial_i\varphi ~,
\nonumber\\
 &\mc{A} = \t{\beta}_1\lambda^{1/2} +\t{\beta}_2\left[\lambda D^i_i
+n^i\xi_{ij}D^j_kn^k\right] +\t{\beta}_4\sqrt{ \frac{\xi}{\gamma}}
\nonumber\\
 & \ \ \ \ \ \ +\t{\beta}_3 \left[ 2\lambda^{1/2}D^{[l}_l
n^{i]}\xi_{ij}D^i_kn^k +\lambda^{3/2}D^{[i}_iD^{j]}_j \right] , \nonumber\\
 & \mc{B} = \t{\beta}_0 +\t{\beta}_1\lambda^{1/2}D^i_j +\t{\beta}_2\lambda
D^{[i}_iD^{j]}_j + \t{\beta}_3 \lambda^{3/2}D^{[i}_iD^j_jD^{k]}_k ~.
\end{align}
One can observe that the Hamiltonian (\ref{hamil}) is not linear in $n^i$ and
hence the equations of motion $\frac{\delta \mc{L}} {\delta n^i} =\mc{C}_i
=0$ do not lead to any constraints on the propagating DoFs, but they
determine $n^i(\gamma_{ij}, \pi^{ij}, \varphi, \pi)$. However, (\ref{hamil})
is now linear in $N$, and thus $\frac{\delta \mc{L}}{N}=\mc{C}=0$ provides
our first constraint on the system.

Consistency requires that $\mc{C}=0$ is valid at all times. If this is
trivially satisfied, then there would be no additional constraints.
Fortunately, we find that the derivative
{\small{
\begin{eqnarray}
\label{constr2}
 \frac{d}{dt}\mc{C}(x)=\{\mc{C}(x),H\} \      \
  \ \ \ \   \ \ \ \   \ \ \ \   \ \ \ \   \ \ \ \ \ \ \ \ \,  \ \ \ \   \ \ \
\
 \ \ \ \   \ \ \ \  \nonumber\\
    =\int d^3y
\left[\{\mc{C}(x),H_0(y)\}-N(y)\{\mc{C}(x),\mc{C}(y)\}\right]
\end{eqnarray}}}
does not vanish trivially, and this indicates that it does imposes the
secondary constraint needed to eliminate the ghost. In the above we have
introduced the Poisson brackets
\begin{align}
 \{\mc{O}_1(x),\mc{O}_2(y)\} &= \int d^3z \left[
\frac{\delta\mc{O}_1(x)}{\delta\gamma_{mn}(z)}
\frac{\delta\mc{O}_2(y)}{\delta \pi^{mn}(z)}\right.\nonumber\\
&\left.   \ \ \ \   \ \ \ \   \ \ \ \ \ \
-\frac{\delta\mc{O}_1(x)}{\delta \pi^{mn}(z)}
\frac{\delta\mc{O}_2(y)}{\delta\gamma_{mn}(z)} \right] \nonumber\\
 &~~+\int d^3z \left[ \frac{\delta\mc{O}_1(x)}{\delta \varphi(z)}
\frac{\delta\mc{O}_2(y)}{\delta \pi(z)}\right.\nonumber\\
&\left.   \ \ \ \   \ \ \ \   \ \ \ \ \ \ - \frac{\delta\mc{O}_1(x)}{\delta
\pi(z)} \frac{\delta\mc{O}_2(y)}{\delta \varphi(z)} \right] ~.
\end{align}
First of all, we need to prove that $\{\mc{C}(x),\mc{C}(y)\}=0$. Otherwise,
the appearance of the lapse $N$ in (\ref{constr2}) would not turn this
consistency requirement into a constraint on the propagating modes as:
$\mc{C}^{(2)}=\int d^3y \{\mc{C}(x),\mc{H}_0(y)\} =0$. Straightforwardly we
calculate
{\small{
\begin{align}
 &\{\mc{C}(x),\mc{C}(y)\} = \{\mc{R}(x),\mc{R}(y)\}\nonumber\\
 & \
+\{\mc{R}_i(x),\mc{R}_j(y)\} D^i_kn^k(x)D^i_l n^l(y)  \nonumber\\
 & \ + \left[ \{\mc{R}(x),\mc{R}_i(y)\} D^i_kn^k(y) +S^{mn}(x)\frac{\delta
\mc{R}_i(y)}{\delta \pi^{mn}(x)}D^i_kn^k(y)\right. \nonumber\\
 & \ \left. - \{\mc{R}(y),\mc{R}_i(x)\}D^i_kn^k(x) +S^{mn}(y)\frac{\delta
\mc{R}_i(x)}{\delta \pi^{mn}(y)}D^i_kn^k(x) \right] ,
\end{align}}}
with
 $S^{mn}(x) = \mc{R}_j\frac{\partial(D^j_rn^r)}{\partial\gamma_{mn}}(x) +
m_g^2 M_p^2 \frac{\partial(\sqrt{\gamma}\mc{B})}{\partial\gamma_{mn}}(x)$.
The involved poisson brackets read
\begin{align}
\{\mc{R}(x),\mc{R}(y)\}&=-\left[\mc{R}^i(x)\partial_{x^i}\delta^{(3)}
(x-y)\right.\nonumber\\
&\left. \ \ \ \ \ \ \ \ -
\mc{R}^i(y)\partial_{y^i}\delta^{(3)}(x-y)\right],\nonumber\\
\{\mc{R}(x),\mc{R}_i(y)\}&=-\mc{R}(y)\partial_{x^i}\delta^{(3)}(x-y),
\nonumber\\
\{\mc{R}_i(x),\mc{R}_j(y)\}&=-\left[\mc{R}_j(x)\partial_{x^i}\delta^{(3)}
(x-y)\right.\nonumber\\
&\left. \ \ \ \ \ \ \ \ -\mc
{R}_i(y)\partial_{y^j}\delta^{(3)}(x-y)\right] .
\end{align}
In order to deal with the delta functions, the smoothing functions $f(x)$ and
$g(x)$ are introduced as follows,
\begin{eqnarray}
 F=\int d^3x f(x)\mc{C}(x) ~,~~ G = \int d^3y g(y) \mc{C}(y) ~,
\end{eqnarray}
and their Poisson bracket can be evaluated as
\begin{eqnarray}
&&\{F,G\}=\int d^3x d^3y f(x) g(x)\{\mc{C}(x),\mc{C}(y)\}\nonumber\\
&&\ \ \ \ \ \ \ \ \ \ =-\int d^3x
(f\partial_ig-g\partial_i f)P^i,
\end{eqnarray}
where
$P^i=(\mc{R}+\mc{R}_jD^j_kn^k)D^i_ln^l+\mc{R}^i+S^{il}\gamma_{il}D^j_kn^k$.
Accordingly, we can simplify the Poisson bracket of ${\mc{C}}$ as
{\small{
\begin{eqnarray}
 \{\mc{C}(x),\mc{C}(y)\} = \left[P^i(x)\partial_{x^i}\delta^{(3)}(x-y)
-P^i(y)\partial_{y^i}\delta^{(3)}(x-y)\right] .\nonumber
\end{eqnarray}
}}
Since we have $P^i =\mc{C}D^i_ln^l+\mc{C}_j\gamma^{ji}
\big|_{\mc{C}=0,~\mc{C}_j =0}=0$, we finally explicitly prove that
\begin{eqnarray}
 \{\mc{C}(x),\mc{C}(y)\}=0 ~.
\end{eqnarray}

In summary, our consistency condition is therefore $\mc{C}^{(2)} =\int
d^3y\{\mc{C}(x), \mc{H}_0(y)\} =0$, where the Poisson bracket in the
integrand writes as
{\small{
\begin{align}
 \{\mc{C}(x),\mc{H}_0(y)\} & = -\{\mc{R}(x),\mc{R}_i(y)\}
(Ln^i+L^i)(y)\nonumber\\
& \ \  \
-D^i_kn^k(x) \{\mc{R}_i(x),\mc{R}_j(y)\}
(Ln^j+L^j)(y) \nonumber\\
& \ \  \ -S^{mn}(x)\frac{\delta \mc{R}_i(y)}{\delta \pi^{mn}(x)} (Ln^i+L^i)
 \nonumber\\
& \ \  \
-\sqrt{\gamma}m_g^2 M_p^2\frac{\partial\mc{B}}{\partial\varphi}
\frac{\partial\mc{R}_i}{\partial\pi} (Ln^i+L^i)\delta^{(3)}(x-y)\nonumber\\
& \ \  \
+L\sqrt{\gamma}m_g^2
M_p^2\left[     \frac{\delta \mc{R}(x)}{\delta \pi^{mn}(y)} A^{mn}(y)\right.
\nonumber\\
& \ \  \ \ \  \ \ \
+ D^i_kn^k(x)\frac{\delta \mc{R}_i(y)}{\delta
\pi^{mn}(x)}A^{mn}(y) \nonumber\\
 & \ \  \ \ \  \ \ \  + \frac{\partial\mc{A}}{\partial\varphi}
\frac{\partial\mc{R}}{\partial\pi} \delta^{(3)}(x-y)
\nonumber\\
& \ \  \ \ \  \ \ \ \left.
+ \frac{\partial\mc{B}}{\partial\varphi}\frac{
\partial\mc{R}_i}{\partial\pi} (D^i_kn^k) \delta^{(3)}(x-y) \right],\nonumber
\end{align}}}
with $A^{mn} = \frac{1}{\sqrt{\gamma}} \frac{\partial
(\sqrt{\gamma}\mc{A})}{\partial\gamma_{mn}}$. After some algebra we finally
obtain
{\small{
\begin{align}
\label{constr21}
 \mc{C}^{(2)} &= \mc{C} \nabla_i(Ln^i+L^i) +m_g^2M_p^2(\gamma_{mn}\pi^k_k
-2\pi_{mn}) A^{mn}\nonumber\\
&  \ \  \
+2L\sqrt{\gamma}m_g^2 M_p^2(\nabla_m A^{mn}) \gamma_{ni}
D^i_kn^k \nonumber\\
&  \ \  \
 + (\mc{R}_jD^i_kn^k-m_g^2M_p^2 \sqrt{\gamma} \gamma_{jk}\bar{\mc{B}}^{ki})
\nabla_i (Ln^j+L^j)
\nonumber\\
&  \ \  \
-m_g^2 M_p^2 \sqrt{\gamma}
\frac{\partial\mc{B}}{\partial\varphi} \partial_i\varphi(Ln^i+L^i)
\nonumber\\
&  \ \  \
-m_g^2
M_p^2 L\frac{\partial\mc {A} }{\partial\varphi} \partial_i\varphi D^i_kn^k ~,
\end{align}}}
where we have defined
{\small{
\begin{align}
 \bar{\mc{B}}^{mn} =& \gamma^{mi}\left\{\t{\beta}_1 \lambda^{-1/2}\xi_{ik}
(D^{-1})^k_j +\t{\beta}_2
\left[\xi_{ik}(D^{-1})^k_jD^l_l-\xi_{ij}\right]\right. \nonumber\\
 &
+ \t{\beta}_3\lambda^{1/2}\left[\xi_{ik}D^k_j -\xi_{ij}D^k_k
\right. \nonumber\\
 &
 \left.\left. \ \ \ \ \ \ \ \ \ \ \ \ \
 +\frac{1}{2}\xi_{ik}(D^{-1})^k_j
 (D^l_l D^k_k-D^l_kD^k_l) \right]\right\}\gamma^{jn}.\nonumber
\end{align}}}

We mention that $\mc{C}$ appears only in the first term of (\ref{constr21}),
and thus this expression does not automatically vanish on the constraint
surface defined by $\mc{C}=0$. Therefore, demanding that $\mc{C}^{(2)}=0$,
imposes the secondary constraint that we needed.

Lastly, a further check needs to be performed, in order to show that no
tertiary constraint exists. This is achieved by
demonstrating that $\{\mc{C}^{(2)}(x),\mc{H}_0\}\ne 0$ and
$\{\mc{C}^{(2)}(x),\mc{C}(y)\}\ne 0$. Therefore the consistency condition,
\begin{eqnarray}
\frac{d}{dt}\mc{C}^{(2)}(x)=\{\mc{C}^{(2)}(x),H\}=0 ~,\nonumber
\end{eqnarray}
gives an equation that determines $N$.

\end{document}